\documentclass[10pt,aps,prd,twocolumn,nofootinbib,showpacs,showkeys,superscriptaddress]{revtex4-2}

\usepackage{psfrag} \usepackage{mathrsfs} \usepackage{amssymb, bm} 
\usepackage{amsmath, amsthm} \usepackage{epstopdf} 
\usepackage[breaklinks=true]{hyperref} \usepackage{enumerate} 
\usepackage{longtable} \usepackage{subfigure} \usepackage{color} 
\usepackage{mathrsfs} \usepackage{graphicx} 
\usepackage{bm,natbib,url,textcase}  \usepackage{xurl} 
\usepackage[title]{appendix}

\usepackage[english]{babel} \usepackage[T1]{fontenc} \usepackage{comment} 
\numberwithin{equation}{section}

\usepackage{geometry}
\allowdisplaybreaks

\hypersetup{
}

\usepackage{tikz}

\newcommand{\be}{\begin{equation}} 
\newcommand{\ee}{\end{equation}} 

\definecolor{purple}{rgb}{1,0,1} \definecolor{lime}{HTML}{a6CE39} 

\newcommand{\orcidicon}{%
	\begin{tikzpicture}
		\draw[lime, fill=lime] (0,0) circle [radius=0.15] 
		node[white] {{\fontfamily{qag}\selectfont \tiny ID}}; 
		\draw[white, fill=white] (-0.0625,0.095) circle 
		[radius=0.007];
	\end{tikzpicture} \hspace{-2mm} }

\newcommand\orcidValerio{{\href{https://orcid.org/0000-0002-2601-1870}{\orcidicon}}}

\begin{document} 
\def\theequation{\arabic{equation}}

\title{Black hole interiors in the thermal view of scalar-tensor gravity}

\author{Valerio Faraoni\orcidValerio} \email[]{vfaraoni@ubishops.ca} 
\affiliation{Department of Physics \& Astronomy, Bishop's University, 2600 
College Street, Sherbrooke, Qu\'ebec, Canada J1M~1Z7}

\begin{abstract}

The thermal view of scalar-tensor gravity is an analogy with a dissipative 
fluid. The scalar degree of freedom excites gravity to a positive 
``temperature'', while Einstein gravity is the ``zero-temperature'' 
equilibrium state. We extend this thermal analogy to the interior region 
near the singularity of spherical, vacuum, uncharged black holes by using 
the universality of the Kasner behaviour near spacelike singularities. The 
singularity is ``hot'', meaning that gravity diverges from general 
relativity. The discussion is then extended to black holes in the presence 
of perfect or imperfect fluids with constant equation of 
state---the latter determines whether 
Einstein gravity is approached or not.

\end{abstract}
\maketitle


\noindent {\it Introduction}---In spite of its success, the standard 
theory of gravity, General 
Relativity (GR) has shortcomings.  It predicts the divergence of physical 
and geometrical quantities and the failure of the spacetime structure 
itself at spacetime singularities inside black holes and in the early 
universe. These singularities are ultimately expected to be cured by 
quantum gravity.  In any case, the lowest-order quantum corrections to GR 
``break'' it by introducing extra degrees of freedom or higher order field 
equations (e.g., 
\cite{Stelle:1976gc,Stelle:1977ry,Callan:1985ia,Fradkin:1985ys,Starobinsky}). 
Furthermore, the standard cosmological model based on GR, the 
$\Lambda$-Cold Dark Matter model, is now subject to various tensions 
\cite{Riess:2019qba,DiValentino:2021izs}. Finally, the current 
acceleration of the cosmic expansion discovered in 1998 with Type Ia 
supernovae was explained with a completely {\em ad hoc} dark energy of 
unknown origin \cite{Amendola:2015ksp}, which is far from satisfactory. 
Cosmologists often prefer the alternative of modifying gravity at 
cosmological scales without invoking dark energy. For this purpose, the 
most well studied and popular alternative to GR is $f(R)$ gravity, where 
the Lagrangian density $f$ is a non-linear function of the Ricci scalar 
$R$ \cite{Sotiriou:2008rp,DeFelice:2010aj,Nojiri:2010wj}.

The simplest modification of GR consists of adding a scalar 
degree of freedom $\phi$, as in scalar-tensor (ST) gravity, which 
corresponds to the inverse of the effective gravitational coupling 
strength $G_\mathrm{eff} \simeq 1/\phi$---the latter becomes a dynamical 
field sourced by matter already in Brans-Dicke gravity 
\cite{Brans:1961sx}, later generalized to what are now called 
``first-generation'' ST theories 
\cite{Bergmann:1968ve,Nordtvedt:1968qs,Wagoner:1970vr,Nordtvedt:1970uv}. 
The class of $f(R)$ theories popular to explain the cosmic 
acceleration without dark energy is a subclass of ST gravity. ST gravity 
with field equations of order not higher than two has been further 
generalized with the rediscovery of Horndeski gravity \cite{Horndeski} and 
its further DHOST extension, which are now the subject of a large 
literature.

First-generation ST gravity 
\cite{Brans:1961sx,Bergmann:1968ve,Nordtvedt:1968qs,Wagoner:1970vr, 
Nordtvedt:1970uv} is described by  the Jordan frame action \be 
S_\mathrm{ST} = \int d^4 x \, \frac{ \sqrt{-g}}{16\pi} \left[ \phi R 
-\frac{\omega(\phi)}{\phi} \, \nabla^c \phi \nabla_c \phi -V(\phi) \right]
+ 
S^\mathrm{(m)} \,, \label{STaction} 
\ee 
where (following the notation of 
Ref.~\cite{Wald:1984rg} and using units in which the speed of light $c$ 
and  Newton's constant $G$ are unity) $g_{ab}$ is  the spacetime metric 
with 
determinant  $g$, covariant 
derivative operator $\nabla_a$,  and   Ricci scalar $R$,   $\omega $ is 
the ``Brans-Dicke coupling''  
(restricted to values $\omega>-3/2$ to avoid the scalar $\phi$ having   
phantom nature and being subject to fatal instabilities), while  
$S^\mathrm{(m)}$ is the matter action.  The Jordan frame field equations 
derived from this action are 
\begin{eqnarray}  
&& R_{ab}-\frac{1}{2} \,  g_{ab} R = \frac{8\pi}{\phi} \, 
T_{ab}^\mathrm{(m)} + T_{ab}^{(\phi)} \,, \label{fe1}\\
&&\nonumber\\
&& \,\,\,T_{ab}^{(\phi)} \equiv \frac{\omega}{\phi^2} \left( \nabla_a \phi
\nabla_b \phi -\frac{1}{2} \, g_{ab} \nabla^c \phi \nabla_c \phi \right) 
\nonumber\\
&& \qquad\quad\quad +\frac{1}{\phi} \left( \nabla_a \nabla_b \phi -
g_{ab} \Box\phi \right) -\frac{V}{2\phi} g_{ab} \, , \, \nonumber\\
&&\nonumber\\
&& (2\omega+3) \Box \phi = 8\pi T^\mathrm{(m)}
+\phi \, V' -2V -\omega' \, \nabla^c\phi \nabla_c\phi \,,\nonumber\\
 \label{fe2} \end{eqnarray} where $R_{ab}$ is the Ricci tensor, 
$T_{ab}^\mathrm{(m)} $ is the matter stress-energy tensor with trace 
$T^{(m)} \equiv g^{ab} T_{ab}^\mathrm{(m)}$, $\Box \equiv g^{ab} \nabla_a 
\nabla_b$, and a prime denotes differentiation with respect to $\phi$.  
$T_{ab}^{(\phi)}$ is an effective stress-energy tensor of $\phi$, which 
originates the thermal view of ST gravity 
\cite{Faraoni:2021lfc,Faraoni:2021jri,Giusti:2021sku,Faraoni:2022gry, 
Giardino:2022sdv,Gallerani:2024gdy}. Provided that the scalar  field 
gradient $\nabla^c\phi$ is timelike and future-oriented, it can be used to 
define the 4-velocity of an effective $\phi$-fluid as \be u^a \equiv 
\frac{\nabla^a \phi}{ \sqrt{-\nabla^c \phi \nabla_c \phi}} 
\,,\label{4-velocity} \ee and $T_{ab}^{(\phi)}$ then assumes the 
well-known dissipative fluid form \be T_{ab} =\rho u_a u_b +P h_{ab} 
+\pi_{ab} + q_a u_b + q_b u_a \,, \label{eq:imperfect} \ee where $\rho$, 
$P$, $\pi_{ab}$ and $q_a$ are, respectively, an effective energy density, 
effective isotropic pressure, effective anisotropic stress tensor, 
effective heat flux density, while $h_{ab} \equiv g_{ab}+u_au_b$ is the 
3-metric on the 3-space seen by timelike observers with 4-velocity $u^a$. 
The effective fluid quantities associated with $T^{(\phi)}_{ab}$, i.e., 
$\rho^{(\phi)}$, $P^{(\phi)}$, $\pi_{ab}^{(\phi)}$ and $q_a^{(\phi)}$, as 
well as its kinematic quantities \cite{Ellis:1971pg}, are reported in 
\cite{Faraoni:2018qdr} for first-generation ST gravity and in 
\cite{Quiros:2019gai,Giusti:2021sku} for its Horndeski generalization.  
Eckart \cite{Eckart:1940te} developed a theory of relativistic dissipative 
fluids which, although non-causal (the heat flux density $q^a$ is 
spacelike) and subject to instabilities, is the first step in the 
description of dissipative fluids and is much simpler than its 
relativistic generalizations \cite{Maartens:1996vi,Andersson:2006nr}. It 
is based on three assumptions in the form of constitutive relations for 
the fluid material, the most important being the generalization of 
Fourier's law for heat conduction \cite{Eckart:1940te} 
\be 
q_a =-{\cal K} 
h_{ab} \left( \nabla^b {\cal T} +{\cal T} \dot{u}^b \right) 
\,,\label{FourierEckart} 
\ee 
where 
${\cal K}$ is the fluid's thermal conductivity, ${\cal T}$ is its 
temperature, and $ \dot{u}^a \equiv u^b \nabla_b u^a$ is its 
4-acceleration. {\em A priori} there is no reason for the effective 
$\phi$-fluid of ST gravity to obey the Eckart constitutive 
relation~(\ref{FourierEckart})  but, 
miraculously, it does: a direct computation of the effective fluid 
quantities and the comparison of $q_a^{(\phi)}$ and $\dot{u}^a$ yield 
\cite{Faraoni:2018qdr,Faraoni:2021lfc,Faraoni:2021jri,Giusti:2021sku}  $ 
q_a^{(\phi)} =-{\cal KT} \dot{u}_a $, from which one 
identifies the product of effective thermal conductivity and effective 
temperature 
\be 
{\cal KT} = \frac{ \sqrt{-\nabla^c\phi \nabla_c\phi} }{ 
8\pi \phi} \,. \label{KTdefinition} 
\ee 
One cannot separate ${\cal K}$ and 
${\cal T}$ and this is the closest that one comes to defining a 
``temperature of gravity'', which is relative to GR. GR, corresponding to 
$\phi =$~const. and ${\cal KT} = 0$, is the ``zero temperature'' state, 
while the excitation of a non-trivial (i.e., propagating) scalar degree of 
freedom $\phi$ corresponds to a thermally excited state at ${\cal KT}>0$.  
The approach of ST gravity to GR (i.e., ${\cal KT} \to 0$), or its 
departure from it (i.e., increasing ${\cal KT}$), are described by the 
equation 
\cite{Faraoni:2018qdr,Faraoni:2021lfc,Faraoni:2021jri,Giusti:2021sku, 
Gallerani:2024gdy} 
\be 
\frac{d\left( {\cal KT}\right)}{d\tau} = 8\pi 
\left( {\cal KT}\right)^2 -\Theta {\cal KT} +\frac{ \Box\phi}{8\pi \phi} 
\,, \label{evolution_general} 
\ee 
where $\tau$ is the proper time along 
the lines of the effective $\phi$-fluid and $\Theta \equiv \nabla_c u^c$ 
is its expansion scalar. Substituting the evolution equation of 
$\phi$~(\ref{fe2}) gives 
\begin{widetext} 
\be 
\frac{d \left( {\cal 
KT}\right)}{d\tau} = 8\pi \left( {\cal KT}\right)^2 -\Theta {\cal KT} + 
\frac{ T^\mathrm{(m)} }{\left( 2\omega + 3 \right) \phi} +\frac{1}{8\pi 
\left( 2\omega + 3 \right)} \left( V' -\frac{2V}{\phi} 
-\frac{\omega'}{\phi} \, \nabla^c\phi \nabla_c\phi \right) \,. 
\label{evolution_general2} 
\ee 
\end{widetext}

Gravity is ``heated'' by positive terms in the right-hand side, which 
contribute to $d\left( {\cal KT}\right)/d\tau>0$ causing departure from 
GR, and is ``cooled'' by negative ones contributing to $d\left( {\cal KT} 
\right)/d\tau<0$ and to gravity approaching GR.

Since $\phi>0$ to keep the effective gravitational coupling 
$G_\mathrm{eff}$ positive, the matter contribution has the sign of 
$T^\mathrm{(m)}$.

Here we consider Brans-Dicke gravity \cite{Brans:1961sx} with 
$\omega=$~const., $V(\phi) \equiv 0$, for which 
Eq.~(\ref{evolution_general2}) reduces to 
\be 
\frac{ d \left({\cal KT} 
\right)}{d\tau} = {\cal KT} \left( 8\pi {\cal KT}-\Theta \right)
+ \frac{ T^\mathrm{(m)} }{\left(2\omega+3 \right) \phi} \,. 
  \label{evolution-reduced} 
\ee

The thermal view of ST gravity is only defined when $\nabla^a \phi$ is 
timelike and future-oriented and is not applicable when $\nabla^a \phi$ 
and $u^a$ are spacelike or null. In first-generation ST gravity, black 
hole no-hair theorems guarantee that, under suitable conditions, the 
spacetime geometry outside the horizon of a stationary, asymptotically 
flat black hole is the same as in GR. The no-hair theorems started with 
Hawking's celebrated theorem establishing that stationary black holes that 
are asymptotically flat in vacuum Brans-Dicke gravity with a massless 
gravitational scalar field $\phi$ are the same as in GR 
\cite{Hawking:1972qk}. Furthermore, the end state of gravitational 
collapse will produce {\em stationary} black holes, which are forced to be 
axisymmetric \cite{Hawking:1971vc}. Over the years, the theorem was 
generalized to include spherical black holes in more general theories 
\cite{Bekenstein:1996pn,Mayo:1996mv,Bhattacharya:2015iha,Faraoni:2017ock} 
and realistic axisymmetric black holes in ST gravities in which 
the scalar field potential $V(\phi)$ has a minimum \cite{Sotiriou:2011dz}. 
Likewise, the violation of no-hair theorems and the search for black hole 
hair in 
alternative theories of gravity is the subject of a large literature (see 
the recent review \cite{Yazadjiev:2025ezx}).

In theories descrived by the action~(\ref{STaction}), provided that 
$V(\phi)$ has a minimum at $\phi_0$ and that $\phi$ and $\omega(\phi)$ are 
not singular outside, or on, the horizon a vacuum stationary black hole in 
ST 
gravity is the same as in GR, i.e., a Kerr black hole because the scalar 
field is forced to have the constant value $\phi_0$ outside the horizon 
and then gravity reduces to GR.  Static, asymptotically flat, spherical 
black holes in vacuum Brans-Dicke gravity are forced to be Schwarzschild, 
with line 
element 
\be 
ds^2=-\left(1-\frac{2m}{r} \right)dt^2 
+\frac{dr^2}{1-2m/r}+r^2 d\Omega_{(2)}^2 \,, 
\ee 
where $d\Omega_{(2)}^2 
\equiv d\vartheta^2 +\sin^2\vartheta \, d\varphi^2 $ is the line element 
on the unit 2-sphere and $m$ is the black hole mass.

What about the {\it interior} of black holes? Here we focus on spherical, 
asymptotically flat black holes. Even though the exterior geometry is 
static Schwarzschild, the interior is dynamic. The coordinates $t$ and $r$ 
switch roles at the horizon $r=2m$ and the interior region is dynamic, 
with time-dependent scalar field $\phi$ and there is a chance to apply the 
thermal view of ST gravity.  We focus on the region near the black hole 
singularity, which is spacelike in Schwarzschild 
\cite{Wald:1984rg,Poisson:2009pwt}. We do not consider rotating or 
electrically charged black holes, for which the singularity can be 
timelike or null (possibly including Cauchy horizons)  
\cite{Wald:1984rg,Poisson:2009pwt}. It is remarkable that spacelike 
singularities in GR exhibit a high degree of universality. In seminal work 
on the approach to spacelike singularities, Belinski, Khalatnikov and 
Lifshitz \cite{Belinsky:1970ew,Belinski:2017fas} have shown that spacelike 
singularities in GR are ultralocal and oscillatory.  In the ultra-local 
limit, the vacuum geometry 
near a spacelike singularity is characterized by an infinite sequence of 
Kasner epochs, in each one of which the geometry is described by a Kasner 
solution of vacuum (i.e., $R_{ab}=0$) GR, with line element 
\be 
ds^2 =- dt^2
+ a_1^2 ( t ) dx^2
+ a_2^2(t) dy^2
+ a_3^2(t) dz^2  \label{kasner} 
\ee 
in Cartesian coordinates, where 
\be a_i(t)= \left( 
  \frac{t}{t_0}\right)^{p_i} \quad\quad (i=1,2,3) \label{1.14} 
\ee 
with constant $t_0$. The constant exponents $p_i$ satisfy 
\begin{eqnarray}
&&p_1 + p_2 + p_3 =1 \,,\\ \nonumber\\
&& p_1^2 + p_2^2 + p_3^2 =1 \,.
\end{eqnarray} 
A unit volume of 3-space in comoving coordinates scales as 
$a^3(t) \equiv  a_1(t)a_2(t) a_3(t) = t/t_0 $. The Kasner solution is a 
special case of the spatially homogeneous and anisotropic Bianchi~I 
geometry which generalizes the spatially flat 
Friedmann-Lema\^itre-Robertson-Walker universe. In the approach to a 
generic spacelike singularity, the exponents $p_i$ depend on the spatial 
coordinates, but this dependence disappears in the ultralocal limit. This 
phenomenology occurs also with certain forms of matter in GR, e.g., for 
perfect fluids with equation of state $P=w\rho$ and $w\leq 1/3$ 
\cite{Ruban75}, for which the singularity is dominated by gravity and 
``matter doesn't matter'', but not for other forms of mass-energy (e.g., 
for perfect fluids with $P>\rho/3$ \cite{Ruban75}).

Let us come to alternative theories of gravity. There are strong 
indications that the Kasner phenomenon persists in modified gravity. The 
Kasner exponents are, in general, functions of the spatial position and 
the geometry reduces to Kasner in the ultralocal limit. In higher-order 
gravity, where multiple energy scales are present due to multiple coupling 
constants in the action, sequences of ``Kasner eons'' have been 
discovered, each one of them consisting of a Kasner epoch with different 
Kasner exponents, with very brief transition periods between one  
epoch and the next. In particular, 
\cite{Deruelle:1989he,Clifton:2006kc,Middleton:2010bv} investigate in this 
regard fourth order gravity, which contains $f(R)$ gravity as a special 
case. This class of theories is equivalent to an $\omega=0$ Brans-Dicke 
theory with $\phi=f'(R) $ and a complicated potential 
\cite{Sotiriou:2008rp,DeFelice:2010aj,Nojiri:2010wj} and it is not 
far-fetched to expect the Kasner phenomenon discovered there 
\cite{Deruelle:1989he,Clifton:2006kc,Bueno:2024qhh} to extend to at least 
first-generation ST gravity with second order equations of motion, and 
possibly to Horndeski \cite{Horndeski} and DHOST gravity. Interesting 
behaviour consisting of sequences of Kasner eons has been found recently 
also in Lovelock gravity black holes \cite{Bueno:2024fzg,Bueno:2024qhh}. 
In first-generation ST gravity, where the exterior (static,  
asymptotically flat) black hole geometry is Schwarzschild, as guaranteed 
by the no-hair theorems, it is reasonable to expect the region near the 
spacelike black hole singularity to have  Kasner geometry. Let us apply 
the  thermal view to this region of ST black hole interiors.

\smallskip
\noindent {\it Black hole interiors via the ST Kasner solution}---The 
Kasner behaviour near the spacelike singularities of ST black holes is 
described by the Kasner solution of ST gravity 
\cite{Ruban:1972bg,Ruban75}, which differs from the GR Kasner solution as 
follows \cite{Ruban:1972bg,Ruban75}. For definiteness, in this section we 
restrict to vacuum Brans-Dicke gravity described by the 
action~(\ref{STaction}) with constant $\omega$ and $V(\phi)=0$, for which 
the Kasner solution is well known, contrary to more general theories.  
The line element is given by Eqs.~(\ref{kasner}), (\ref{1.14}), while 
$\phi=\phi(t)$ because of spatial homogeneity. In terms of the average 
scale factor $
a(t) \equiv \Big[ a_1(t) a_2(t) a_3(t) \Big]^{1/3} $ 
and of the individual directional Hubble functions $ H_i(t) \equiv 
\dot{a}_i / a_i $, the field equations for the Kasner 
geometry~(\ref{kasner}) reduce to \cite{Ruban:1972bg,Ruban75} 
\begin{eqnarray}
&& \frac{1}{a^3} \, \frac{d}{dt} \left( a^3 H_i
\right) +H_i \, \frac{\dot{\phi}}{\phi}=0 \,,\\
&&\nonumber\\
&& \frac{d}{dt} \left( a^3 \dot{\phi} \right)=0 \,,\\ 
&& \nonumber\\ 
&&H_1 H_2  + H_2 H_3 + H_1 H_3 = \frac{\omega}{2} \left( 
\frac{\dot{\phi}}{\phi} 
   \right)^2 +\frac{ \ddot{\phi}}{\phi} \,, 
\end{eqnarray} 
where an  overdot denotes differentiation with respect to the comoving 
time $t$. They admit the first integrals 
\begin{eqnarray}
&& \dot{\phi} = \frac{C_1}{a^3} \,,\label{firstintegral1}\\ 
&&\nonumber\\ 
&& \frac{d (a^3)}{dt} = \frac{C_2}{\phi} \,,
\end{eqnarray} 
with $C_{1,2}$ integration constants. The 
solution  is \cite{Ruban:1972bg,Ruban75} 
\begin{eqnarray} 
a_i(t) & &= \left( \frac{t}{t_0} 
\right)^{\frac{p_i}{1+C}} \,,\\
&&\nonumber\\
\phi(t) &=& \phi_0 \left( \frac{t}{t_0} \right)^{ \frac{C}{1+C}} \,, \quad 
\quad C =\frac{C_1}{C_2} \,. \label{sol-phi} 
\end{eqnarray} 
Again, 
$p_1+p_2+p_3=1$ but now $a^3 = \left(t/t_0 \right)^{ \frac{1}{C+1} }$ and 
\cite{Ruban:1972bg,Ruban75} 
\be 
p_1^2 + p_2^2 + p_3^2 =1 - C\left(\omega C 
-2 \right) \,, 
\ee 
which marks the difference with GR. The ST Kasner 
solution reduces to the GR one for $C=0$. It was already noted in 
\cite{Ruban75} that the scalar degree of freedom alters substantially the 
GR behavior at the initial singularity. The 
thermal view interprets this situation as the divergence of gravity from 
GR, with ${\cal KT}$ going infinitely far away from zero.

In order for the timelike vector field $\nabla^a \phi$ to be 
future-oriented it must be 
\be 
\dot{\phi} = \frac{C}{C+1} \, 
\frac{\phi_0}{t_0} \left( \frac{t_0}{t} \right)^{\frac{1}{C+1} } <0 \,; 
\ee 
since $\phi_0>0$, it must be $ \frac{C}{C+1}<0$, which is satisfied in 
the range 
\be
-1<C<0 \,. 
\ee 
Outside of this range of values of $C$, the thermal view of 
 ST gravity is not applicable. For these values of $C$, the 3-volume $ 
 a^3  \to 0 $, while $\phi  \to +\infty$ and 
 $G_\mathrm{eff}\to 0$ at the singularity $t\to 0^{+}$.

To quantify the departure from GR near the spacelike singularity, note 
that 
\be 
{\cal KT} = \frac{ |\dot{\phi}|}{8\pi \phi} = \frac{|C|}{8\pi \left( 
C+1 \right) t} \,, 
\ee 
diverges as $1/t$ at the singularity $t\to 0^{+}$. 
By contrast, the GR Kasner solution corresponding to $C=0$ has constant 
$\phi$ and ${\cal KT}=0$ and the time-dependence of $a_i(t)$ and 
$\phi(t)$, although still power-law, is quite different from the GR case, 
as noted in \cite{Ruban:1972bg,Ruban75}. Hence, in the absence of matter 
and scalar field potential, Brans-Dicke gravity diverges away from GR near 
spacelike singularities, confirming an insight proposed in 
\cite{Faraoni:2021lfc,Faraoni:2021jri,Giusti:2021sku}: gravity is ``hot'' 
near these singularities. Although the discussion is simple, the 
universality of the Kasner behaviour near spacelike singularities is quite 
powerful, allowing general conclusions about the departure of gravity from 
GR in these regions.

Ref.~\cite{Ruban75} also reports the special degenerate solution 
for $C=-1$ in the parameter  range $-3/2<\omega<
-4/3$: 
\begin{eqnarray} 
a_i(t) & = & a_i^{(0)} \, \mbox{e}^{ p_i \, t/t_0 }  \,, \label{exp1}\\
&&\nonumber\\
\phi (t) & = & \phi_0 \, \mbox{e}^{ -\, t/t_0 } \,,\label{exp2} 
\end{eqnarray} 
for which $\dot{\phi}=-\phi /t_0 <0 $ and $\nabla^a \phi$ is 
future-oriented. This solution is reported also in \cite{Ruban:1972bg} 
with the incorrect sign in the argument of the exponential in $\phi(t)$, 
but 
it is straightworward to check that~(\ref{exp1}) is correct using 
Eq.~({\ref{firstintegral1}) with $C=-1$. For this solution,  
\be 
{\cal KT} = 
\frac{| \dot{\phi}|}{8\pi \phi} = \frac{1}{8\pi t_0} 
\ee 
and this solution 
always remains different from the ($C=0$)  GR solution, but at a finite 
constant 
``distance'' ${\cal KT}$ from it, which seems bizarre. To understand this 
situation, note that the expansion scalar 
\be 
\Theta =\frac{1}{a^3} \, 
\frac{d(a^3)}{dt} = \frac{1}{t_0} =8\pi {\cal KT} \ee and, since we are 
{\em in vacuo}, \be \frac{d \left( {\cal KT}\right)}{dt}={\cal KT}\left( 
8\pi {\cal KT}-\Theta\right)=0 \,, 
\ee 
allowing ${\cal KT}$ to remain 
constant. Indeed, Ref.~\cite{Faraoni:2025alq} identified the $\left( 
\Theta, {\cal KT} \right) $ plane as a convenient description of vacuum ST 
gravity and 
the line $8\pi {\cal KT}=\Theta$ as a line of fixed points that cannot be 
crossed dynamically. For expanding 3-spaces with 
$\Theta >0$, this line separates the region $8\pi {\cal KT}>\Theta$, in 
which $d\left( {\cal KT}\right)/d\tau>0$ and gravity always departs from 
GR, from the other region $d\left( {\cal KT}\right)/d\tau<0$ in which ST 
gravity always converges to GR \cite{Faraoni:2025alq}.   The fine-tuned  
solution~(\ref{exp1}), (\ref{exp2}) lies on this special line, along which 
$8\pi {\cal KT}=\Theta=$~const. 
 
\smallskip
\noindent {\it Including matter, $V(\phi)$ and $\omega(\phi)$}---Let us 
focus now on Eq.~(\ref{evolution-reduced}) including matter, where 
the proper time 
$\tau$ along the effective fluid lines coincides with the Kasner comoving 
time $t$ because 
\be 
u^0 \equiv \frac{ g^{0b}\partial_b \phi }{ 
\sqrt{-\nabla^c\phi \nabla_c \phi} } = \frac{ -\dot{\phi} }{|\dot{\phi}|} 
= 1= \frac{dt}{d\tau} \,. 
\ee 
Furthermore, assume that matter is described by a 
perfect or imperfect fluid with linear barotropic equation of state 
$P=w\rho $ with $w=$~const., then $ T^\mathrm{(m)}=-\rho + 3P = \left( 
3w-1 \right)\rho $.  A radiation fluid with $P=\rho/3$ (or any form of 
conformal matter, which has $T^\mathrm{(m)}= 0$)  does not affect ${\cal 
KT}$. An imperfect (real) fluid has the form~(\ref{eq:imperfect}) and, as 
for perfect fluids, its trace is $T^\mathrm{(m)}=-\rho+3P$.

Let us estimate the powers with which the three terms in the right-hand 
side of Eq.~(\ref{evolution-reduced}) scale as $t \to 0^{+}$.  
Following the authors of \cite{Ruban75}, who discuss the field equations 
instead of Eq.~(\ref{evolution-reduced}), we use the vacuum solution to 
estimate the powers in the gravitational degrees of freedom. This 
procedure is justified by the fact that the fluid density and pressure 
scale the same way with $a$ for a test fluid or for a gravitating one 
and is corroborated by the exact Bianchi~I solutions for 
gravitating matter \cite{Ruban75}.  

In the ST vacuum Kasner solution we 
have ${\cal KT} \sim 1/t$, $\Theta \sim 1/t $, while $\rho \sim P = w\rho 
\sim a^{-3(w+1)} \sim t^{-\, \frac{w+1}{C+1} } $, therefore 
$  T^\mathrm{(m)} / \phi \sim  1/ t^{ \frac{w+C+1}{C+1} } $.  
The gravity terms $\left({\cal KT}\right)^2 \sim t^{-2} $ and 
  ~$-\Theta {\cal KT} \sim t^{-2}$ dominate over the matter term $ 
  \frac{T^\mathrm{(m)}}{ \left(2\omega+3\right)\phi}$ as $ t\to 0^{+}$ if 
\be 
\frac{w+C+1}{C+1}< 2 
\ee 
which (since $C+1>0$) is equivalent to $w<C+1$. On the contrary, the 
matter term 
$$ 
\frac{T^\mathrm{(m)} 
  }{\left( 2\omega+3\right) \phi} = \frac{ (3w-1)\rho_0 \, t_0^{ \frac{ 
  w+C+1}{C+1} } }{\left( 2\omega+3\right) \phi_0 \, t^{ \frac{ w+C+1}{C+1} 
  }
} 
$$ 
dominates if $w>C+1$. The same conclusion is reached in \cite{Ruban75} by 
comparing terms in the field equations and using the vacuum Kasner 
solution to estimate the gravitational and matter terms (cf. Eq.~(15) of 
\cite{Ruban75}).

Gravity and the matter term scale with the same power of $1/t$ as $t\to 
0^{+}$ if $w=C+1 >0$, in which case (not discussed in \cite{Ruban75} or 
subsequent literature) we have 
\be 
\frac{ d\left({\cal KT}\right)}{dt} =\left[ \frac{C}{8\pi (C+1)}
 + \frac{ \left( 3C+2 \right) \rho_0 t_0^2}{\phi_0} \right] \frac{1}{t^2} 
   \,; 
\ee
since $ -1< 3C+2< 2$, the sign of the right-hand side depends on the 
initial conditions, which determine $\rho_0, \phi_0, t_0$, and on $w=C+1$. 
If $-1<C<-2/3$, one can in principle tune the initial condition so that 
${\cal KT}=$~const. (similarly to the line $8\pi {\cal KT}=\Theta$ for 
vacuum), although such fine-tuning would be physically 
uninteresting.

In addition to matter, one can think of including in the picture a scalar 
field potential $V(\phi)$, which complicates the analysis significantly. 
If $V$ is quadratic, it disappears from Eq.~(\ref{fe2}) for $\Box\phi$ and 
from Eq.~(\ref{evolution-reduced})  for ${\cal KT}$ because it only 
enters these equations  through the combination $\phi V'-2V $. Otherwise, 
$V(\phi)$ makes 
Eq.~(\ref{fe2}) nonlinear and difficult to solve. As noted in 
\cite{Faraoni:2025alq} using Eq.~(\ref{evolution_general2}), if $V(\phi)$ 
grows with $\phi$ faster than $\phi^2$, it ``heats up'' gravity, while it 
``cools'' it if it grows slower than $\phi^2$. Likewise, if present, the 
last term in Eq.~(\ref{evolution_general2}), $\frac{ \omega' \dot{\phi}^2 
}{8\pi \left( 2\omega+3 \right) \phi}$, contributes to ``heating gravity'' 
if $\omega(\phi)$ is monotonically increasing. However, without an 
explicit solution one cannot estimate the size of the terms in the 
right-hand side of Eq.~(\ref{evolution-reduced}) and decide which ones are 
dominant.

\smallskip
\noindent {\it Conclusions}---We have extended the thermal view of ST 
gravity to cover the interior 
region of ST black holes near their spacelike singularities, thus filling 
an obvious gap in the applications of this new formalism.  The discussion 
of the 
previous section appears simple, perhaps too simple, but it is a virtue of 
the new thermal view of ST gravity that it simplifies the description of 
ST physics and offers a unified frame to interpret technical results in a 
coherent way. The universality of the Kasner behaviour near spacelike 
singularities allows one to reach quite general conclusions about the 
departure of gravity from GR in these regions, where the spacetime 
structure fails. A first implication is that, if quantization adds a 
scalar degree of freedom to gravity, as in $f(R)$ gravity 
\cite{Sotiriou:2008rp,DeFelice:2010aj,Nojiri:2010wj} or in Starobinski 
inflation \cite{Starobinsky}, then one should not attempt to quantize the 
Einstein equations when looking for Kasner transitions in 
singularity-free bouncing universes (which are of great current interest), 
but one should address 
the field equations of ST gravity instead  \cite{deCesare:2019suk}, 
or adopt an approach not tied 
to the classical field equations. The conclusion may even be more radical: 
if the thermal view is taken seriously, gravity and its field equations 
could be emergent and not suitable for quantization as proposed, e.g.,  in 
Jacobson's thermodynamics of spacetime 
\cite{Jacobson:1995ab,Eling:2006aw}. The latter is quite similar in 
spirit, although completely different in substance, to the thermal view of 
ST gravity.

We have not said anything about timelike singularities such as those 
encountered in electrically charged black holes 
\cite{Wald:1984rg,Poisson:2009pwt} or about null singularities (such as 
thunderbolt singularities \cite{Hawking:1992ti}) or Cauchy horizons.  
Although the general idea that gravity is ``hot'' (i.e., deviates from GR) 
near all spacetime singularities presumably remains valid 
\cite{Faraoni:2021lfc,Faraoni:2021jri,Giusti:2021sku}, no generic 
behaviour like the Kasner phenomenology is known near these other types of 
singularities. Explicit examples aiming to falsify or support the general 
idea that such spacetime singularities also imply a catastrophic 
divergence from GR will be sought for elsewhere.

Singularities are still present in scalar-tensor
gravity and the thermal formalism, being purely classical, does nothing to
cure them. All that it offers is the realization
of how these spacetime singularities are different from those in GR and
why---a cure for these 
singularities will have to come from somewhere else.
The understanding of ``different'' singularities proposed here may have
some impact on certain approaches to quantization, but there is no reason 
to believe this impact to be crucial.

The main point of the new thermal view of scalar-tensor gravity is 
that it offers a new framework to understand apparently unrelated 
phenomena that have appeared in the literature over three decades. They 
are indeed related and the new thermal formalism, which is rather concise, 
unifies apparently disconnected phenomena in a single view, simplifying 
their description. However, all these aspects of scalar-tensor gravity 
need to be revisited one by one in the new picture and then the unifying 
character of the latter is not manifest. We refer the reader to 
\cite{Faraoni:2021lfc,Faraoni:2021jri,Giusti:2021sku,Faraoni:2022gry, 
Giardino:2022sdv,Gallerani:2024gdy} for details and we will extend the 
study to new phenomena in future work.

\smallskip
\noindent This work is supported, in part, by the Natural Sciences \& 
Engineering Research Council of Canada (grant 2023-03234).


\end{document}